\documentclass[aps,prb,a4paper,twocolumn]{revtex4-1} 
\pdfoutput=1 

\usepackage{bm}
\usepackage{amsmath}
\usepackage{amsfonts}
\usepackage{amssymb}
\usepackage{verbatim} 

\usepackage[normalem]{ulem}
\usepackage{xcolor} 
\usepackage{graphicx} 

\usepackage{float}            
\usepackage{times}
\usepackage{mathptmx} 

\usepackage{hyperref}
\hypersetup{
pdfborder={0 0 0}, 
colorlinks=true, 
linkcolor=blue,
citecolor=blue,
urlcolor=blue
}

\newif\iftwopi
\twopitrue

\iftwopi
  \newcommand{\newomega}{\omega}
\else
  \newcommand{\newomega}{\nu}
\fi

\newcommand{\corer}{a} 
\newcommand{\cylr}{R}
\newcommand{\rea}{\mathrm{Re}_{\alpha}}

\newcommand{\praim}{'}
\newcommand{\praimpraim}{''}

\newcommand{\alphap}{\alpha\praim}

\newcommand{\critical}{\mathrm{c}}
\newcommand{\tc}{T_\critical}
\newcommand{\ttc}{T/T_\critical}

\newcommand{\Omegac}{\Omega_\mathrm{s}}
\newcommand{\newomegac}{\newomega_\mathrm{s}}

\newcommand{\rv}{\mathbf{r}}
\newcommand{\sv}{\mathbf{s}}

\newcommand{\svh}{\mathbf{\hat{s}}} 
\newcommand{\svp}{\sv\praim}
\newcommand{\svhp}{\svh\praim}
\newcommand{\svpp}{\sv\praimpraim}
\newcommand{\Omegav}{\boldsymbol{\Omega}}

\newcommand{\velocity}{v}
\newcommand{\velocityv}{\mathbf{\velocity}}

\newcommand{\vov}{\velocityv_{\mathrm{\omega}}}
\newcommand{\vlv}{\velocityv_{\mathrm{L}}}
\newcommand{\vnv}{\velocityv_{\mathrm{n}}}
\newcommand{\vsv}{\velocityv_{\mathrm{s}}}

\newcommand{\vbv}{\velocityv_{\mathrm{b}}}

\newcommand{\counterflow}{_\mathrm{cf}}
\newcommand{\vvcf}{\velocityv\counterflow}

\newcommand{\OstermeierGlaberson}{\mathrm{OG}}
\newcommand{\vcog}{\velocity_{\critical,\OstermeierGlaberson}}

\newcommand{\Omegacbl}{\Omega_\mathrm{m}}
\newcommand{\newomegacbl}{\newomega_\mathrm{m}}

\newcommand{\vlz}{\velocity_{\mathrm{L}z}}
\newcommand{\vlfi}{\velocity_{\mathrm{L}\phi}}
\newcommand{\vfz}{\velocity_{\mathrm{F}z}}
\newcommand{\vffi}{\velocity_{\mathrm{F}\phi}}

\newcommand{\hethree}{${^3}$He}

\newcommand{\hethreeb}{${^3}$He-B}

\newcommand{\hethreebs}{{${^3}$He-B }}

\newcommand{\evhz}{\mathbf{\hat{e}}_{z}}
\newcommand{\evhr}{\mathbf{\hat{e}}_{r}}
\newcommand{\evhfi}{\mathbf{\hat{e}}_{\phi}}

\newcommand{\comments}[1]{}

\begin{document}

\title{Asymptotic motion of a single vortex in a rotating cylinder}

\author{J.M. Karim\"aki$^{*,\dagger}$} 
\author{R. H\"anninen$^{*}$} 
\author{E.V. Thuneberg$^{\dagger}$}
\affiliation{$^*$O.V. Lounasmaa Laboratory, Aalto University, P.O.\,Box 15100, FI-00076 AALTO, Finland}
\affiliation{$^\dagger$Department of Physics, University of Oulu, Oulu, Finland}

\date{April 8, 2013}

\begin{abstract}
We study numerically the behavior of a single quantized vortex  in a rotating cylinder. We study in 
particular the spiraling motion of a vortex in a cylinder that is parallel to the rotation axis.
We determine the asymptotic form of the vortex and its axial and azimuthal propagation velocities
under a wide range of parameters. We also study the stability of the vortex line and the effect of 
tilting the cylinder from the rotation axis.
\end{abstract}

\maketitle

\section{Introduction}\label{s.intro}

Since the discovery of quantized vortices, the motion of those vortices under various conditions
has attracted continued attention of researchers 
\cite{RP195517,PhysRev.138.A429,schwarz85,sonin87,DonnellyBook,tsubotaetal00,barenghidonnellyvinen01}.
In recent years emphasis has been shifting to the study of the so-called quantum turbulence and 
numerical simulations with a large number of vortices
\cite{vinenniemela02,kivotidesandleonard03,tsubotaetal03,tsubotaetal04,finneetal06b,eltsovetal09,baggaleyandbarenghi11}.
This requires a considerable amount of computing power, especially when calculations are performed 
using the full Biot-Savart equations \cite{schwarz85}. There is, however, still a need to better 
understand the motion of a single vortex. Since the motion of a curved vortex line is somewhat
counter-intuitive, and solving the equations analytically is difficult and prone to errors, it is 
convenient to use numerical simulations. For single vortices, it also becomes a realistic possibility 
to scan a large volume of parameter space, i.e.~various combinations of pressure $p$, temperature $T$, 
rotational velocity $\Omega$, vessel size and shape, and initial vortex configurations.

\begin{figure}[!tb]
\centerline{
\includegraphics[width=0.49\linewidth]{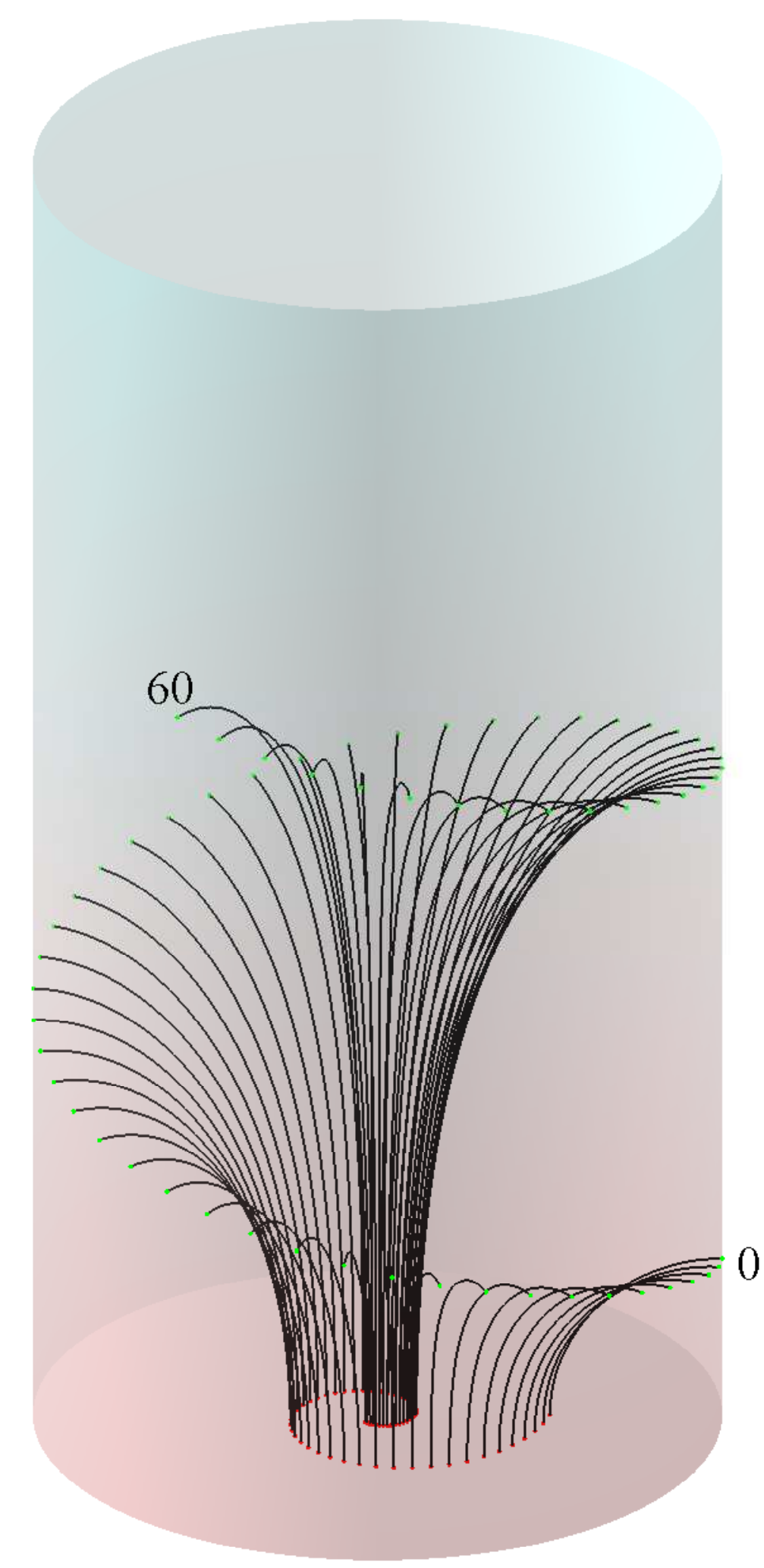} 
}
\caption{(Color online) The evolution of a vortex in a rotating coordinate system. The total time span 
$2.79 / \alpha \Omega$ comprises 61 snapshots. Other parameters are $\rea = 3.63$ ($\sim 0.4\tc$ in \hethreeb) and 
\iftwopi
     $2\pi\cylr^2\Omega/ \kappa = 85.5$.
\else
     $\cylr^2\Omega/ \kappa = 13.6$.
\fi
\label{fig:Evolution-of-a-vortex}}
\end{figure}

\begin{figure}[!tb]
{a)\vspace{7mm}}\includegraphics[width=0.65\linewidth]{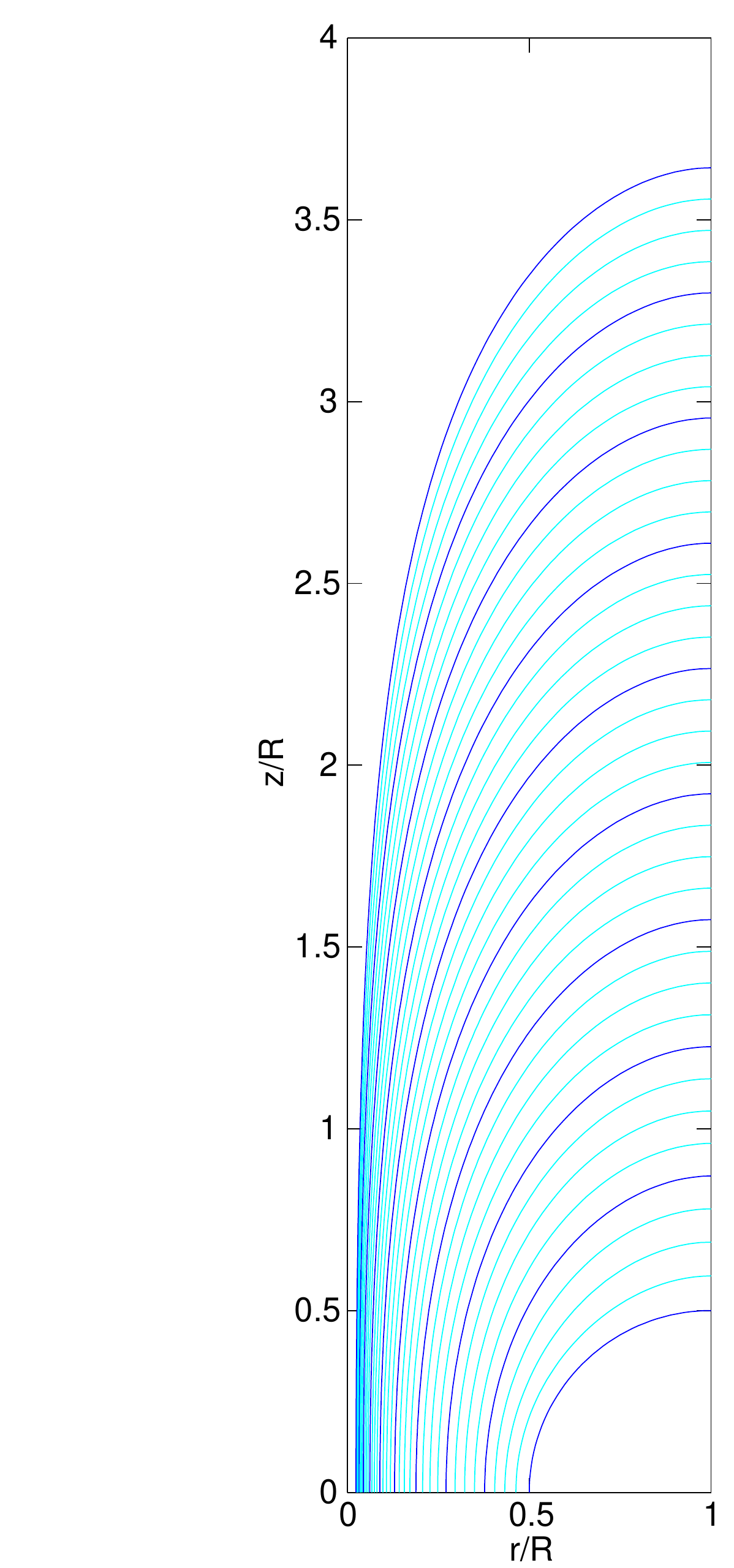}\\
{b)\vspace{0mm}}\includegraphics[width=0.83\linewidth]{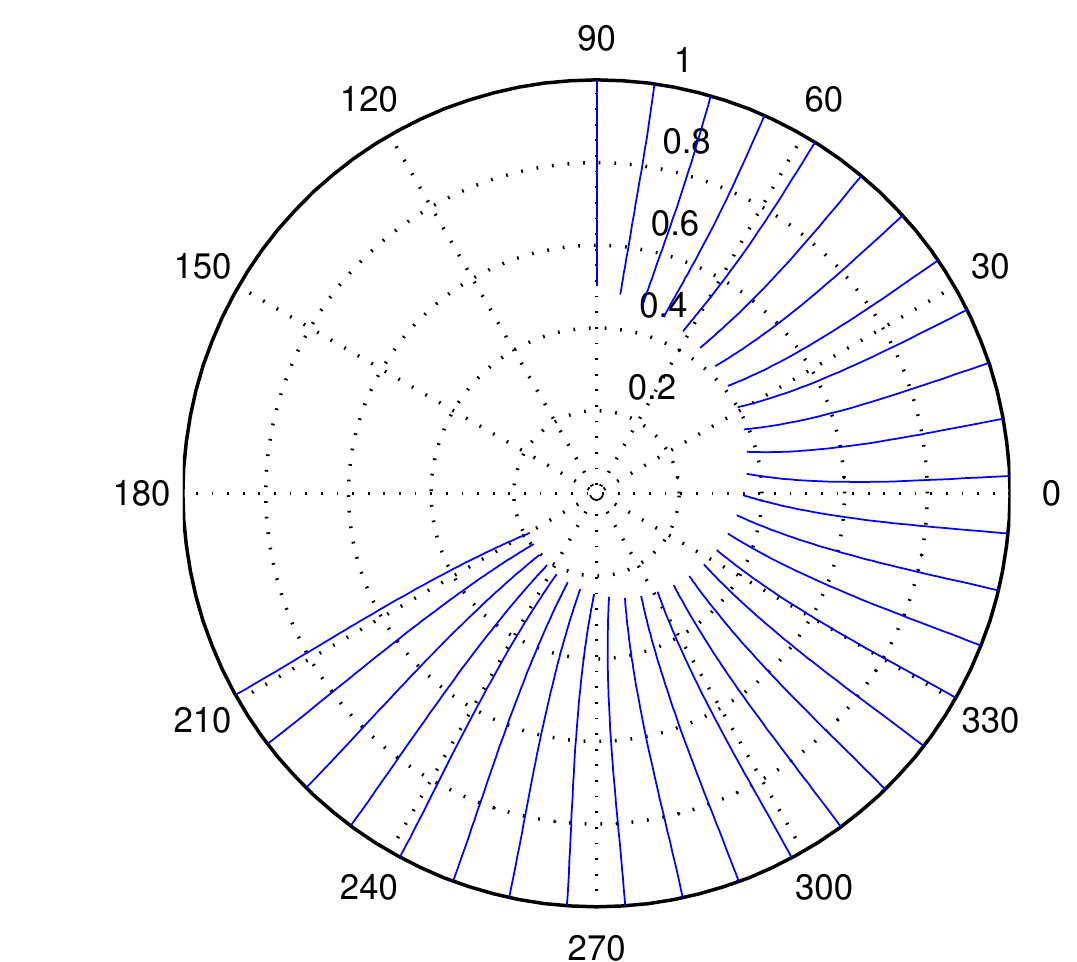}
\caption{(Color online) The evolution of the vortex shown in Fig.\:\ref{fig:Evolution-of-a-vortex} 
presented in cylindrical coordinates fixed to the rotating frame.
a) $(r,z)$ plot with 37 snapshots in time span $4.19 / \alpha\Omega$. 
b) Polar plot with 31 snapshots in time span $1.40 / \alpha\Omega$. 
\label{fig:Evolution-of-a-vortex-2}}
\end{figure}

An existing computer software, previously used mainly for studying the large scale behavior of many
vortices \cite{hanninenetal05,eltsovetal06,hanninenetal09}, is applied to study the motion of a single 
vortex line. More specifically, we study the motion of a single superfluid vortex filament
in a rotating infinite cylinder, as illustrated in Figs.~\ref{fig:Evolution-of-a-vortex} and 
\ref{fig:Evolution-of-a-vortex-2}. This case has recently been studied using analytic approximations 
in Ref.~\onlinecite{sonin11}. A closely related problem, where a wire is placed on the axis of the cylinder,
has been studied earlier\cite{zieveetal92,misirpashaevandvolovik92,schwarz93,sonin94}.

Our main point of interest is the asymptotic velocity of the vortex end touching the side wall of 
the cylinder. We calculate the evolution of the vortex from the initial state using the vortex filament 
model of the two-fluid paradigm with Biot-Savart formalism, until the asymptotic situation is reached.
We assume the normal fluid component of the velocity to be in rigid body rotation. Although we have 
used \hethreeb-specific parameters in our calculations, the methods and the results can be generalized 
to vortices in other superfluids and Bose-Einstein condensates in many cases.

One surprising effect discovered in our calculations is the stability of the vortex, even in the low 
temperature limit. This contradicts the expectations based especially on experimental results in 
\hethreeb \cite{finneetal03,eltsovetal09}. We assume that the discrepancy with experiments is caused by 
some surface effects not accounted for in our model (such as pinning), or due to some uncontrolled heat 
leaks or superflows that may destabilize the vortices in the experiments .

\section{Model}\label{s.model}

We study a superfluid in an infinite cylinder of radius $\cylr$ that rotates at constant angular 
velocity $\Omegav$. For the most part in this paper the axis of rotation is assumed to be the cylinder 
axis. The effects caused by a rotation axis that is tilted with respect to the cylinder axis are 
considered in Secs.~\ref{s.stability} and \ref{s.tilt} at the end of the paper. We use cylindrical 
coordinates $(r,\phi,z)$ fixed to the cylinder. Our study is based on the two-fluid model, where 
the normal and superfluid components have velocities $\vnv$ and $\vsv$, respectively. The normal 
component is assumed to be in rigid body rotation, $\vnv = \Omegav \times \rv$. This is a good 
assumption in \hethreeb, because mutual friction is  weak in comparison to the viscosity of the 
normal fluid \cite{sonin87}.

The superfluid velocity $\vsv$ is determined by vortex lines\cite{schwarz85}. The positions on the 
vortex lines  are given by $\sv(\xi,t)$, which is parametrized by the vortex length $\xi$ and 
time $t$. (The direction in which $\xi$ increases conforms to the right-hand rule for the superfluid 
circulation around the vortex.) Partial derivatives with respect to $\xi$ are denoted by primes.
Then, the unit tangent of the vortex core line is the first derivative of $\sv$ with respect to $\xi$. 
This is denoted by $\svp$ (or $\svhp$ to emphasize that it is a unit vector). The superfluid velocity 
$\vsv$ is calculated using the (full) Biot-Savart formalism:
\begin{equation}
\vsv(\rv,t) = \vov + \vbv, 
\label{eq:v biot-savart}
\end{equation}
with
\begin{equation}
\vov
= \frac{\kappa}{4\pi} \int_{\mathcal{L}}\frac{(\sv_{1}-\rv) \times d\sv_{1}}{|\sv_{1}-\rv|^{3}},
\label{eq:v biot-savart-1}
\end{equation}
where $\mathcal{L}$ denotes all the vortex lines and $\kappa$ is the circulation of the superfluid 
velocity around a vortex line; all the positions and velocities are in laboratory coordinates.
The first term, $\vov$, in Eq.\:(\ref{eq:v biot-savart}) needs to be numerically calculated from 
the integral (\ref{eq:v biot-savart-1}). It has a divergence on the vortex core, which has to be 
cut off at the vortex core radius
\footnote{We use the core radius parameter $\corer$ and the corresponding cut-off procedure
in the sense that the velocity of a vortex ring in our calculations equals (within numerical error)
the velocity of a vortex ring with a hollow core in a classical ideal fluid. See 
Ref.~\onlinecite{DonnellyBook}, pages 22-23 for how different core models relate to the velocity 
of a classical vortex ring.
}
$\corer$, see Refs.~\onlinecite{schwarz85} and \onlinecite{hanninen09} for details. The second term, 
$\vbv$, is the boundary field, or image velocity field, needed to prevent flow across the vessel 
boundary. The boundary field $\vbv$ can be calculated from the Laplace equation, or by the method of 
image vortices. To calculate $\vbv$ we use the method of approximate image vortices as in 
Ref.~\onlinecite{hanninen09}.

The derivative of $\sv(\xi,t)$ with respect to time defines the vortex line velocity $\vlv$ except that 
the component of $\vlv$ parallel to the line is not defined. The equation of motion is commonly written
in laboratory coordinates as\cite{schwarz85}:
\begin{equation}
\vlv = \vsv + \alpha \svhp \times (\vnv-\vsv) - \alphap \svhp \times [\svhp \times (\vnv-\vsv)]. 
\label{eq:equmotion1}
\end{equation}
It contains the mutual friction parameters $\alpha$ and $\alphap$. They depend on temperature and pressure, 
but the temperature dependence is usually much stronger, e.g., in the case of superfluid \hethree-B
the temperature dependence is dramatic \cite{bevanetal97,hall99}. A useful new quantity involving 
$\alpha$ and $\alphap$ can be defined as $\rea := (1-\alphap) / \alpha$. It has some similarity with the 
Reynolds number defined in normal fluids \cite{finneetal03,volovik03,kopnin04}. In \hethreebs the 
quantity $\rea(T)$ decreases monotonically from $\infty$ to $0$, when the temperature increases from $0$ 
to $\tc$. Furthermore, the experimentally accessible range of $\rea$ in \hethreebs essentially
covers this whole range. Table \ref{t.table} contains some approximative values of the parameters
$\alpha$, $\alphap$, and $\rea$, as functions of temperature.

Using the fact that the normal component is in rigid body motion ($\vnv=\Omegav \times \rv$ in the 
laboratory coordinates), we can rewrite the equation of motion (\ref{eq:equmotion1}) in rotating coordinates 
(where $\vnv=0$) as:
\begin{equation}
\vlv = \alpha(\rea \vsv - \svhp \times \vsv).
\label{eq: equmotion2} 
\end{equation}
From now on all velocities will be in rotating coordinates, unless explicitly stated otherwise.

We study in particular the velocity of the vortex end in the long-time limit ($t \rightarrow \infty$). 
For that we  define  $\vlz$ and $\vlfi$ by writing in the rotating frame:
\begin{eqnarray}
\vlv = \vlz \evhz + \vlfi \evhfi.
\label{e.tervel}\end{eqnarray}
It is also possible to define the climbing angle $\beta$ of the vortex in the rotating frame 
by $\tan\beta = \vlz / \vlfi$. The  climbing angle in the laboratory frame satisfies 
$\tan\beta = \vlz /(\vlfi+ \cylr\Omega )$.

\begin{table}[tbh]
\begin{center}
 \begin{tabular}{|c||c|c|c|}
\hline
$T/T_\mathrm{c}$ &  $\alpha$      & $\alpha'$      & $\mathrm{Re}_{\alpha}$  \\  \hline
\hline
0.00    &  0             &  0             &  $\infty$      \\  \hline
0.25    &  1.265828E-02  &  5.255042E-03  &  7.858453E+01  \\  \hline
0.30    &  4.701086E-02  &  3.032361E-02  &  2.062665E+01  \\  \hline
0.35    &  1.200093E-01  &  8.150743E-02  &  7.653512E+00  \\  \hline
0.40    &  2.328964E-01  &  1.556440E-01  &  3.625457E+00  \\  \hline
0.45    &  3.467192E-01  &  2.429385E-01  &  2.183500E+00  \\  \hline
0.50    &  4.565632E-01  &  3.333683E-01  &  1.460108E+00  \\  \hline
0.55    &  5.620771E-01  &  4.201522E-01  &  1.031616E+00  \\  \hline
0.60    &  6.732573E-01  &  4.999895E-01  &  7.426737E-01  \\  \hline
0.65    &  8.104478E-01  &  5.719796E-01  &  5.281283E-01  \\  \hline
0.70    &  1.004341E+00  &  6.365370E-01  &  3.618920E-01  \\  \hline
0.75    &  1.295975E+00  &  6.946998E-01  &  2.355757E-01  \\  \hline
0.80    &  1.736739E+00  &  7.478107E-01  &  1.452085E-01  \\  \hline
0.85    &  2.448636E+00  &  7.974992E-01  &  8.269943E-02  \\  \hline
1.00    & $\infty$       &  1             &  0             \\  \hline
\end{tabular}
\end{center}  
\caption{Tabulated values of $\alpha$, $\alphap$, and $\rea$ used in the numerical calculations.
These values approximately correspond to those in $^3$He-B~at $29.34$ bar at the temperatures shown 
on the left. The limit values of the parameters at $T=0$ and $T=\tc$ are also shown.
\label{t.table}}\end{table}

\section{Scaling properties}\label{s.scaling}

The parameter space of the system (without considering the parameters describing the initial vortex configuration) 
consists of mutual friction parameters $\alpha$ and $\alphap$, the vortex core radius $\corer$, the circulation 
quantum $\kappa$, the cylinder radius $\cylr$, the angular velocity $\Omega$, and for a tilted cylinder the tilting 
angle $\theta$. The exploration of this parameter space is simplified by dimensional analysis. Let us, for example, 
study the asymptotic axial velocity $\vlz$ (\ref{e.tervel}) of the vortex end in the absence of tilting, $\theta = 0$. 
According to dimensional analysis, the dimensionless velocity $\vlz / \cylr\Omega$ can only depend on dimensionless 
quantities appearing in the problem, which in this case are $\cylr / \corer$, $\cylr^2\Omega / \kappa$, $\alpha$, 
and $\alphap$. Thus the parameter space is four-dimensional. Using properties specific to the present problem, we 
show in the following that this space can further be reduced to three dimensions exactly and to two dimensions 
approximately. 
 
Let us consider the evolution of a vortex state from a fixed initial configuration. We see from 
Eq.\:(\ref{eq: equmotion2}) that varying $\alpha$ but keeping $\rea$ constant has the effect of changing the time 
scale only.
\footnote{We would like to point out that this scaling property is independent of the shape of the container.
The scaling property fails, when the field $\vbv$ is not constant in time, unless its speed of time variation 
is also scaled by the factor $\alpha$.}
This means that the dimensionless time is $\alpha\Omega t$ and the dimensionless line velocity is $\vlv / \alpha\cylr\Omega$, 
but otherwise there is no dependence on $\alpha$. (Note that the dimensionless superfluid velocity 
$\vsv / \cylr\Omega$ does not have $\alpha$.) Since the asymptotic velocity is obtained in the limit 
$t \rightarrow \infty$, the dependence on $\alpha\Omega t$ drops out. Since the system is dissipative 
(for $\alpha \not = 0$), the same final state can be obtained by a variety of initial states, i.e.~the details 
of the initial state are not important either. Thus we conclude that the asymptotic velocity has the form
\begin{equation}
\vlz = \alpha\cylr\Omega \, G_z\!\!\left(\frac{\cylr^2\Omega}{\kappa},\frac{\cylr}{\corer},\frac{1-\alphap}{\alpha}\right)\!,
\label{e.vscaling-z}
\end{equation}
with some, as yet unknown, function $G_z$. A similar analysis can be done for the azimuthal velocity,
and the result can be written as
\begin{equation}
\label{e.vscaling-phi}
\vlfi = (\alphap-1)\cylr\Omega \,G_{\phi}\!\!\left(\frac{\cylr^2\Omega}{\kappa},\frac{\cylr}{\corer},\frac{1-\alphap}{\alpha}\right)
\end{equation}
with a function $G_{\phi}$. Note that Eq.\:(\ref{e.vscaling-phi}) is valid only in the rotating frame; the laboratory 
frame azimuthal velocity $\vlfi + \cylr\Omega$ does not have this form. Because the parameter $\cylr^2\Omega/\kappa$ 
appears repeatedly, we define a dimensionless
\iftwopi
     angular velocity:
     \begin{eqnarray}
     \newomega := \frac{2\pi\cylr^2\Omega}{\kappa}.
    \label{e.omega} \end{eqnarray}
\else
     rotational velocity:
     \begin{eqnarray}
     \newomega := \frac{\cylr^2\Omega}{\kappa}.
      \label{e.omega}\end{eqnarray}
\fi
We also remind that the last parameter in Eqs.\ (\ref{e.vscaling-z}) and (\ref{e.vscaling-phi}) has the short-hand 
$\rea \equiv(1-\alphap) / \alpha$.

\begin{figure}[tb]
\includegraphics[width=0.6\linewidth]{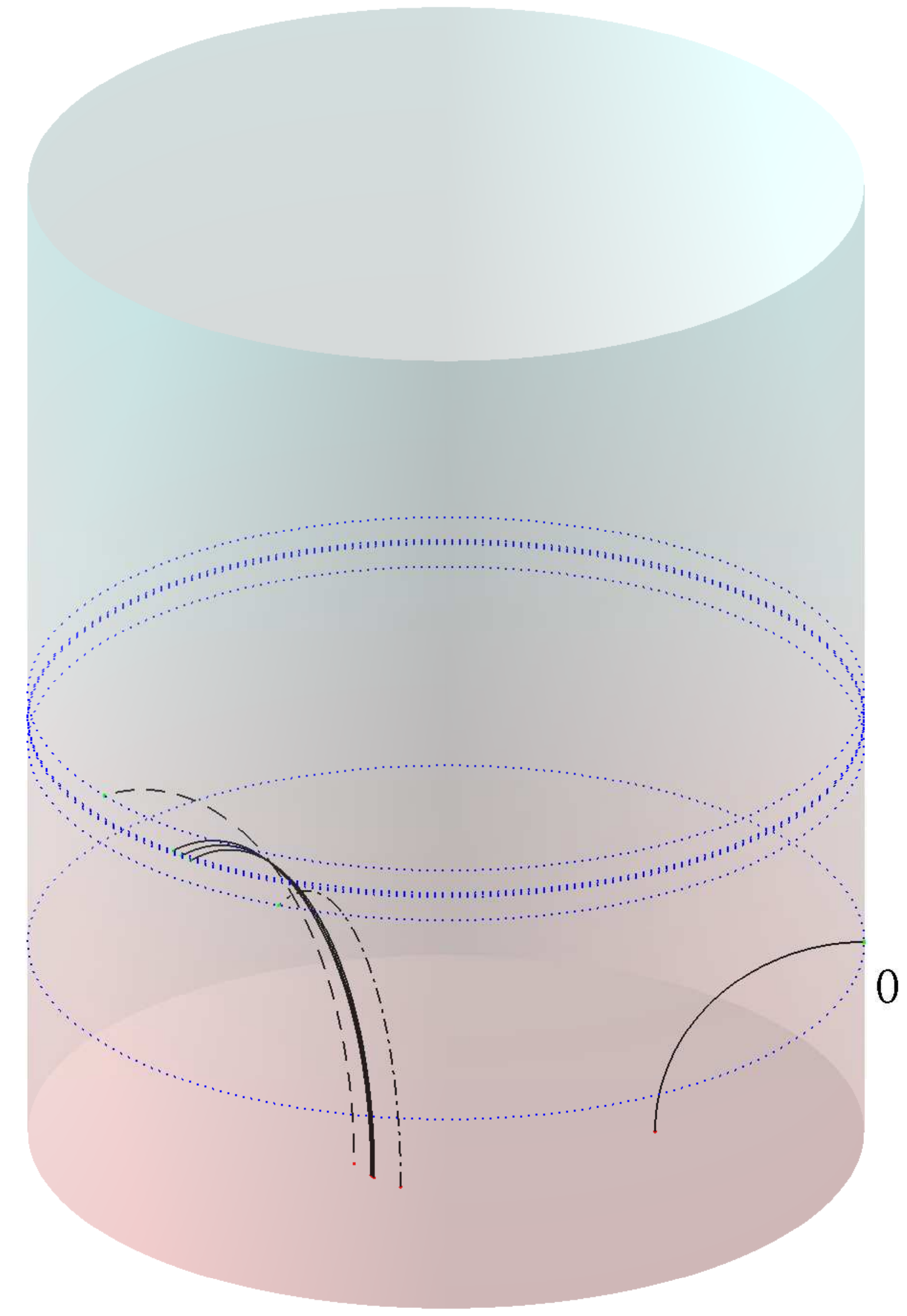} 
\caption{(Color online) The effect of the parameter $\cylr / \corer$. The solid lines have $\cylr / \corer$ 
equal to 1, 0.5, and 0.25 times $1.77 \times10^5$ while other parameters have constant values
\iftwopi
     ($\alpha \Omega t = 0.75$, $\rea = 3.63$, and $\newomega = 85.5$).
     These can be contrasted to two cases, which differ by $\pm10\%$ in the value
     of $\Omega$ from the middle case (the dashed line has $\alpha\Omega t = 0.82$ and 
     $\newomega = 94.1$ and the dash-dotted line has $\alpha\Omega t = 0.67$ and $\newomega = 77.0$).
\else
     ($\alpha \Omega t = 0.75$, $\rea = 3.63$, and $\newomega = 13.6$).
     These can be contrasted to two cases, which differ by $\pm10\%$ in the value
     of $\Omega$ from the middle case (the dashed line has $\alpha\Omega t = 0.82$ and 
     $\newomega = 15.0$ and the dash-dotted line has $\alpha\Omega t = 0.67$ and $\newomega = 12.2$).
\fi
The vortices are shown in the rotating coordinate system.
\label{fig:Scaling-properties-demonstrated-2.}}
\end{figure}

Another, in this case approximate, simplification is offered by the fact that $\vlv$ has only a weak logarithmic 
dependence on $\cylr / \corer$, see e.g., Refs.~\onlinecite{swansonanddonnelly85} and \onlinecite{schwarz88}.
Hence, different values of $\cylr / \corer$ have only a minor effect on the the results; this is demonstrated 
in Fig.\:\ref{fig:Scaling-properties-demonstrated-2.}. 
Because of this weak dependence, we have fixed $\cylr / \corer = 1.77 \times10^5$ in all other calculations in 
this paper except in the ones presented in Fig.\:\ref{fig:Front-Comparison}. We consider explicitly the case, 
where the lower end of the  cylinder (small $z$ side) is in a vortex state with positive circulation
and the angular velocity $\Omega$ is positive. The case of negative circulation and negative angular velocity
has the same $\vlz$ and opposite $\vlfi$.

\section{Critical angular velocities}\label{s.criticalv}

Two critical angular velocities are important for our problem. They can be found analytically by studying the 
following expression of free energy $F$ (per unit length) of the superfluid with an axially oriented vortex 
line displaced at the distance $r$ from the cylinder axis:\cite{PhysRev.138.A429,PhysRev.161.189,campbellziff79}
\begin{equation}\label{eq: free energy}
F(r,\Omega)\! =\! \frac{\rho_s\kappa^2}{4\pi} \left[ 
\ln\!\frac{\cylr}{\corer}
\!+\!\ln\! \left(1\!-\!\frac{r^2}{\cylr^2}\right)\!-\!\frac{2\pi \cylr^2\Omega}{\kappa} \! \left(1\!-\!\frac{r^2}{\cylr^2}\right) \right]. 
\end{equation}
This expression is valid when $\corer \ll  \cylr$.

In the order of increasing $\Omega$, the first critical velocity $\Omegacbl$ corresponds to an axial vortex 
becoming metastable. It can be obtained from the condition:
\begin{equation}
\left.\frac{\partial^2 F(r,\Omegacbl)}{\partial r^2}\right|_{r=0} = 0.
\label{eq: bean-livingston diff eq}
\end{equation}
($\partial F / \partial r = 0$ is automatically satisfied at $r=0$ for all $\Omega \geq 0$.) This condition gives:
\begin{equation}
\Omegacbl = \frac{\kappa}{2\pi\cylr^2}.
\end{equation}
In dimensionless form: 
\iftwopi
     $\newomegacbl \equiv (2\pi\cylr^2 / \kappa) \Omegacbl= 1$.
\else
     $\newomegacbl \equiv (\cylr^2 / \kappa) \Omegacbl= 1/2\pi$.
\fi
This first critical angular velocity is analogous to a critical field in a type II superconductor,
above which the so-called Bean-Livingston barrier\cite{PhysRevLett.12.14} is created.

The second critical velocity $\Omegac$ corresponds to a single axial vortex becoming absolutely stable, 
i.e.~its free energy (8) becomes less than the Landau-state free energy $F=0$. This leads to:
\begin{equation}
\Omegac = \frac{\kappa}{2\pi\cylr^2} \ln\!\frac{\cylr}{\corer}.
\label{eq:Oc2}
\end{equation}
In dimensionless form: 
\iftwopi 
     $\newomegac \equiv 2\pi\cylr^2\Omegac / \kappa = \ln(\cylr / \corer)$,
     which in our case ($\cylr / \corer = 1.77 \times10^5$) has the value 12.08.
\else
     $\newomegac \equiv \cylr^2\Omegac / \kappa = (1/2\pi) \ln(\cylr / \corer)$,
     which in our case ($\cylr / \corer = 1.77 \times10^5$) has the value 1.92.
\fi
The second critical angular velocity is an analog of the lower critical magnetic field, at which the phase 
transition between the Meissner state and the mixed state takes place in type II superconductors.

The one-vortex state loses absolute stability when the two-vortex state becomes stable\cite{PhysRev.161.189,campbellziff79},
but it still remains metastable. The critical velocity $\velocity_\critical$ for vortex nucleation at the cylindrical 
wall limits the metastability of the one-vortex state to $\newomega<\newomega_\critical := 2\pi \cylr \velocity_\critical/\kappa$.
We study below a wide range of $\newomega$, but the results are applicable only as long as $\newomega$ remains 
below $\newomega_\critical$. For $^3$He-B in a vessel with smooth walls $\newomega_\critical \gg 1$ can be achieved,
which justifies the present study.

\section{Asymptotic vortex velocity in a non-tilted cylinder}\label{s.asymp}

The evolution of the vortex is solved by numerically integrating Eq.\:(\ref{eq: equmotion2}) forward in time in the 
rotating frame. Our numerical scheme uses a half-infinite cylinder, but the results for the asymptotic velocity and 
the vortex form in an untilted cylinder are equally valid for an infinite cylinder. For large rotation velocities 
we typically use a quarter vortex ring as our initial condition. The radius of the ring is half of the cylinder radius. 
The ring is situated so that the vortex starts from the bottom of the cylinder and ends at the cylinder wall.
For small rotational velocities ($\Omega \lesssim \Omegac$) the initial configuration was taken from previous iterations 
at somewhat larger $\Omega$. This was necessary in order to reach the steady state quickly enough, or even in order to 
avoid the shrinking away of the initial configuration. Both ends of the vortex are allowed to move freely. Any vortex end 
touching the wall is always normal to it. In principle, a vortex line may also form a closed loop, without touching the 
boundary at all. A vortex loop may be created when a vortex line reconnects with itself. If a vortex loop comes very close 
to the wall, it reconnects with it, that is, snaps open, with the two vortex ends connecting to the wall orthogonally.

The use of a quarter vortex ring as the initial condition is not necessary. What matters is whether the required 
asymptotic evolution (if it exists) is reached from the initial condition or not, and a quarter ring is suitable in 
a large parameter range. However, if the temperature is low and the shape of the vortex is far from the asymptotic 
one, the vortex may oscillate a long time before reaching a stable form. For $\rea \gtrsim 100$,
even the quarter ring vortex leads to long-lasting oscillations in the vortex shape.

Our goal is to study the velocity of the end point of the vortex $\vlv$ (\ref{e.tervel}). Before going to the general 
case, we consider some limiting cases where analytic solutions are found.

In the limit $\newomega \gg 1$,  the cylinder wall at $r = \cylr$ can be thought of as a plane and we assume that 
the vortex line approaches the wall as a straight line normal to it. We can now put $\Omegav = \Omega \evhz$,
$\rv = \cylr \evhr$, a point on the large cylinder wall (or plane) and $\svhp = \evhr$. This gives 
$\vsv = -\cylr\Omega \evhfi$ in rotating coordinates, and inserting it into Eq.\:(\ref{eq: equmotion2}) we get:
\begin{equation}
\vlv = \cylr\Omega  \left[ \alpha \evhz + (\alphap - 1) \evhfi \right].
\label{eq:limiting case cyl rot}
\end{equation}
Hence $\vlz = \alpha\cylr\Omega$, $\vlfi = (\alphap-1)\cylr\Omega$, and $\tan \beta = -\rea^{-1}$.

Another exact result corresponds to vanishing velocity of the vortex line, $\vlv = 0$, at the critical angular 
velocity $\Omega = \Omegac$ (\ref{eq:Oc2}). Since at $\Omega = \Omegac$ the one-vortex and no-vortex states are in 
equilibrium, there is no force to drive the vortex. Therefore at $\Omega=\Omegac$ the vortex must be stationary 
in the rotating frame, for all temperatures $0< T < \tc$.

A third special case is the limit $T\rightarrow 0$. In this limit the normal fluid component vanishes and thus the 
vortex does not feel the rotation of the container. Therefore the shape and the motion of the vortex, as measured 
in the laboratory coordinates, do not depend on the rotational velocity of the vessel. As the previous limiting 
case ($\Omega = \Omegac$) extends to this limit as well, we expect that the shape of the vortex in the 
$T\rightarrow 0$ limit is the same as the eqilibrium shape at $\Omega = \Omegac$. This implies that the velocity 
in the laboratory system is $\vlz =0$ and $\vlfi = \cylr\Omegac$. In the rotating frame these translate to 
$\vlz =0$ and $\vlfi= \cylr (\Omegac - \Omega)$.

\begin{figure}[tb]
\includegraphics[clip,width=0.95\linewidth]{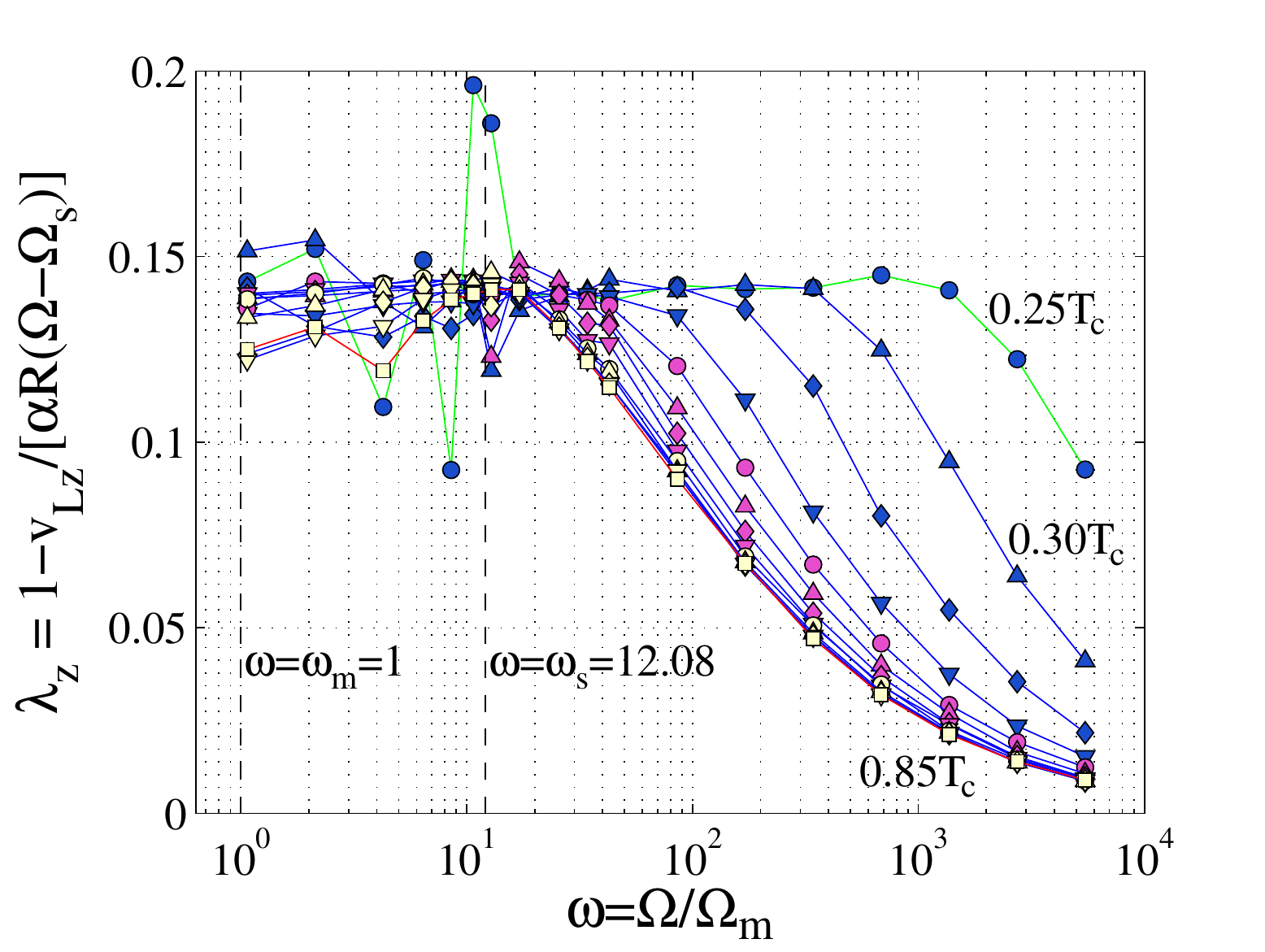}
\caption{(Color online) The axial velocity of the vortex  $\vlz$ expressed by  $\lambda_z$ (\ref{e.vlzp}). 
The curves are plotted as a function of
\iftwopi
     $\newomega \equiv 2\pi\cylr^2\Omega / \kappa$.
\else
     $\newomega \equiv \cylr^2\Omega / \kappa$.
\fi
The used values of $\rea$ are given in Table \ref{t.table}. They correspond approximately to temperatures
$\ttc = 0.25,0.30,0.35,...,0.85$ in \hethreeb. The two vertical dashed lines correspond to the two critical 
angular velocities $ \Omegacbl$ and $\Omegac$.
The value of $\Omegac$ used in calculating $\lambda_z$ and $\lambda_\phi$ is an average obtained numerically 
from the calculated data: $\Omegac = 12.55\Omegacbl$, which deviates slightly from the theoretical value
$\Omegac = \ln(\cylr/\corer)\Omegacbl = 12.08\Omegacbl$. Using this numerical value instead of the theoretical 
one eliminates a peak in $\lambda_z$ and $\lambda_\phi$ near $\Omegac$ that we believe to be a numerical artifact.
The data still shows some numerical error, especially at low temperatures and for angular velocities
$\Omega \approx \Omegac$. The numerical calculation is consistent with the expectation that  $\lambda_z$
is a constant at $\Omega = \Omegac$, and gives the value $\lambda_z \approx 0.14$.
\label{fig:lambda-z}}
\end{figure}

Based on these limiting cases, we can refine the dependences (\ref{e.vscaling-z}) and (\ref{e.vscaling-phi}) by introducing
\begin{align}
\vlz&= \alpha \cylr \left(\Omega - \Omegac \right)\left(1-\lambda_z\right),\label{e.vlzp}\\
\vlfi&= (\alpha'-1) \cylr \left(\Omega - \Omegac \right)\left(1-\lambda_\phi\right)\label{e.vlfip}.
\end{align}
We have defined new dimensionless functions $\lambda_z$ and $\lambda_\phi$, which, similarly as $G_z$ and $G_\phi$, 
are functions of $\omega$ (\ref{e.omega}), $R/a$, and $\rea$. The rationale here is that the $\lambda$ functions 
are small compared to unity, so that a crude approximation can be obtained by neglecting $\lambda_z$ and 
$\lambda_\phi$ in Eqs.~(\ref{e.vlzp}) and (\ref{e.vlfip}). This already implies that the sign of $\vlz$ is determined 
by $\Omega - \Omegac$: the vortex grows for $\Omega>\Omegac$ and shrinks for $\Omega<\Omegac$. The sign of $\vlfi$ is 
determined similarly, but noting that the coefficient $\alphap-1$ is negative.
\begin{figure}[htb]
\includegraphics[clip,width=0.95\linewidth]{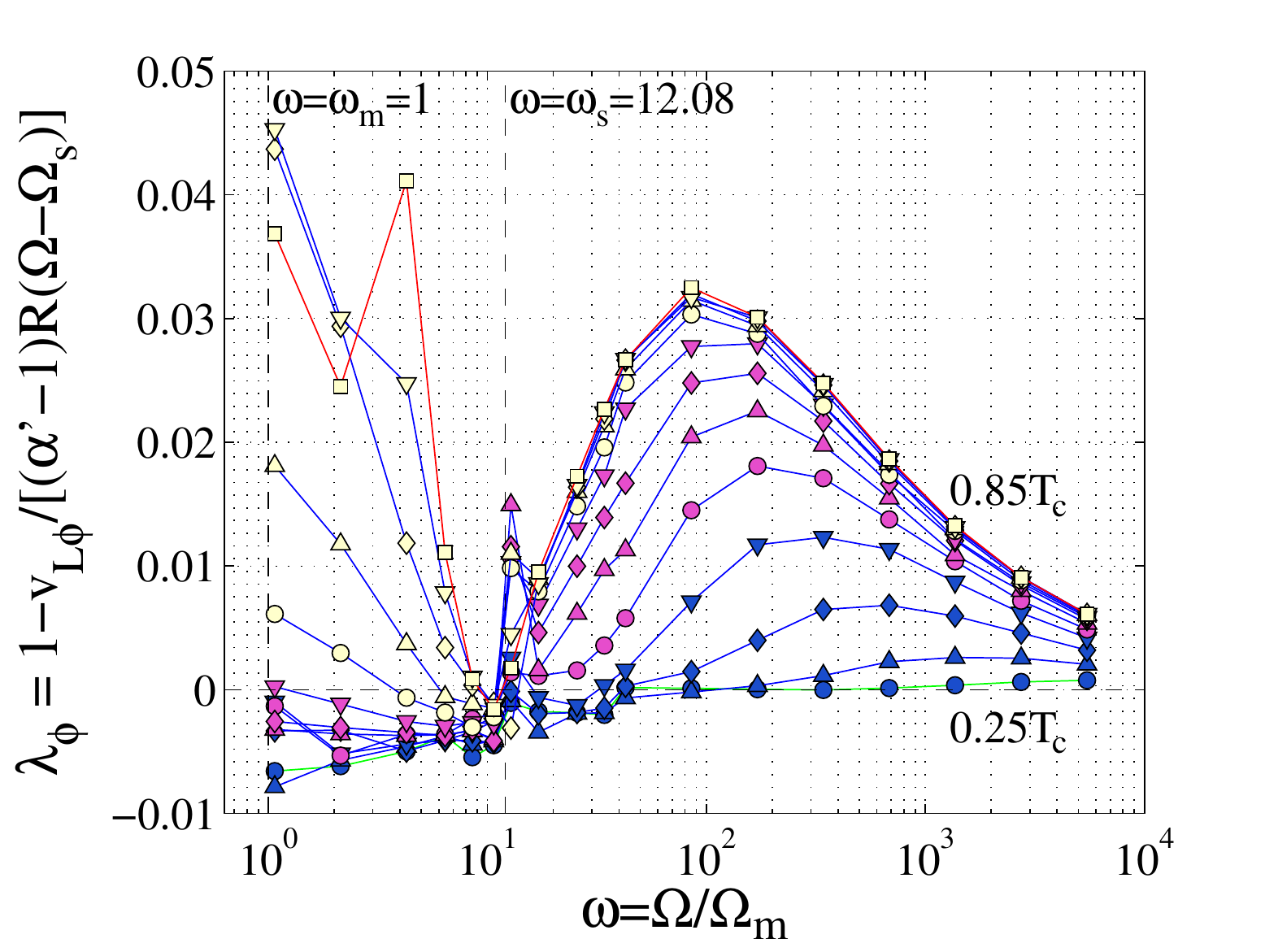}
\caption{(Color online) The azimuthal velocity of the vortex  $\vlfi$ in the rotating frame, expressed by 
$\lambda_\phi$ (\ref{e.vlfip}). The curves are plotted as a function of $\newomega$ at the same values of the 
parameter $\rea$ as used in Fig.\:\ref{fig:lambda-z}, again corresponding to approximate temperatures from 
$0.25\tc$ to $0.85\tc$ in \hethreeb. The two vertical dashed lines correspond to the two critical angular 
velocities $\Omegacbl$ and $ \Omegac$. The value of $\Omegac$ used for calculating $\lambda_\phi$ is the same as 
for $\lambda_z$ in Fig.~\ref{fig:lambda-z}. Here also, numerical errors are prominent for low temperatures and 
for $\Omega \lesssim \Omegac$. The numerical results are consistent with the expectation that $\lambda_\phi$ vanishes 
at $\Omega = \Omegac$, and in the low temperature limit. The negative values of $\lambda_\phi$ most likely arise 
from numerical inaccuracy.  
\label{fig:lambda-phi}}
\end{figure}

The numerically calculated asymptotic velocities $\vlz$ and $\vlfi$ are shown in Figs.~\ref{fig:lambda-z} and 
\ref{fig:lambda-phi}. The results are presented using the $\lambda$ functions defined in Eqs.~(\ref{e.vlzp}) and 
(\ref{e.vlfip}). The figures show some scatter especially for $\Omega$ close to $\Omegac$ that arises from inaccuracies 
in the numerical calculation. In spite of this, we can conclude that the limiting cases mentioned above are consistent 
with the data. In particular, both $\lambda_z$ and $\lambda_\phi$ are finite functions, which guarantees stationarity 
at $\Omega=\Omegac$. Both $\lambda_z$ and $\lambda_\phi$ approach zero in the limit of large $\omega$. In the limit 
$T\rightarrow 0$, $\lambda_\phi$ approaches zero irrespective of $\omega$. There is no restriction on $\lambda_z$ in 
the limit $T\rightarrow 0$ since the prefactor $\alpha\rightarrow 0$ in Eq.\:(\ref{e.vlzp}). 

From Figs.\ \ref{fig:lambda-z} and \ref{fig:lambda-phi} one can see that both $\lambda_z$ and $\lambda_\phi$ are constants (independent 
of $T$) at $\Omega=\Omegac$. This can be understood by considering the mutual friction as a small perturbation in the 
vicinity of $\Omega=\Omegac$. As the mutual friction parameters appear explicitly in Eqs.\ (\ref{e.vlzp}) and 
(\ref{e.vlfip}), the functions $\lambda_z$ and  $\lambda_\phi$ should not depend on $(1-\alphap)/\alpha$ at $\Omega=\Omegac$.
Generalizing this argument, we can say that around $\Omega=\Omegac$ there is an {\em inertial regime}, where the 
vortex shape is dominated by non-dissipative forces. This regime is characterized by $\lambda_z\approx 0.14$. There 
is a complementary {\em dissipative regime}, where $\lambda_z\ll 0.14$. The boundary between the two is at 
$\Omega-\Omegac\approx \rea  \Omegac$. Both the inertial and dissipative limits seem to correspond to vanishing 
$\lambda_\phi$, but non-vanishing values appear in the cross-over regime.

We note that although the curves in Figs.~\ref{fig:lambda-z} and \ref{fig:lambda-phi} are labeled by temperature,
the results are completely general. The temperatures correspond to different values of $\rea$ according to table 
\ref{t.table} and the dependence on $\alpha$ or $\alphap$ is through the scaling relations (\ref{e.vscaling-z}) 
and (\ref{e.vscaling-phi}), or (\ref{e.vlzp}) and (\ref{e.vlfip}). 

For large $\omega$ we can identify the approximate power laws
\begin{equation}
\begin{split}
\vlz  &\simeq \alpha\cylr\Omega [ 1-\left(\Omega/\Omega_{z}\right)^{k_{z}} ],\\
\vlfi &\simeq (\alphap-1)\cylr\Omega [ 1-\left(\Omega/\Omega_{\phi}\right)^{k_{\phi}} ].
\end{split}
\label{eq:v_Lz}
\end{equation}
Fitting in the range $\Omega \in [85\Omegacbl, 5472\Omegacbl]$ at $\rea = 0.083$ ($\sim 0.85\tc$) gives 
$k_{z}   = -0.72$ and $ \Omega_{z} = 10.6\Omegacbl$.
Fitting in the range $\Omega \in [684\Omegacbl, 5472\Omegacbl]$ at $\rea = 0.083$ ($\sim 0.85\tc$) gives
$k_{\phi} = -0.71$ and $\Omega_{\phi} = 6.47\Omegacbl$.
\begin{figure}[H]
\noindent \begin{centering}
\iftwopi
     \includegraphics[width=0.9\linewidth]{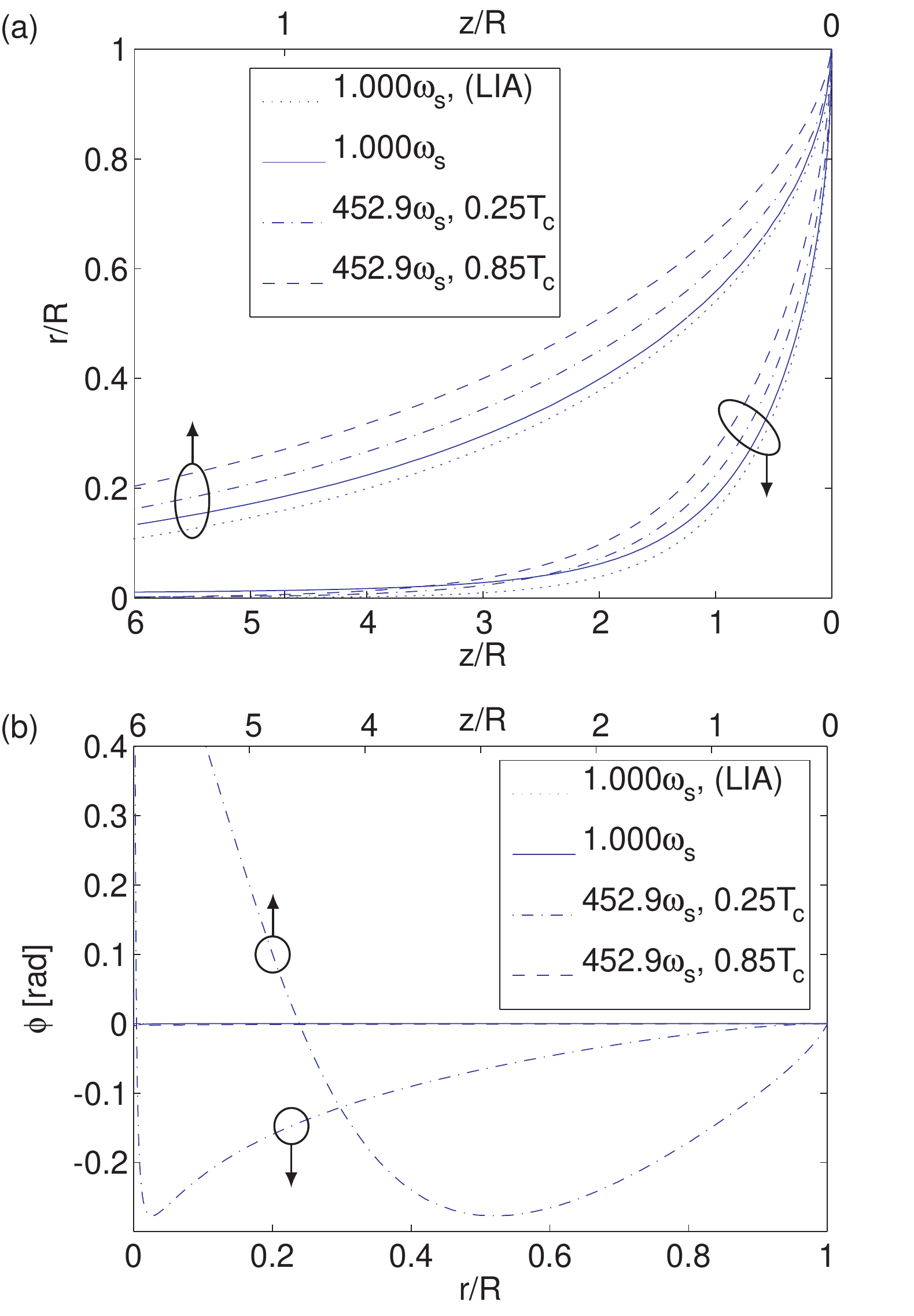}
\else
     {a)\hspace{2.3mm}\vspace{2mm}}\includegraphics[scale = 0.53]{am11_fig6a}\\
     {b)\hspace{2.3mm}\vspace{2mm}}\includegraphics[scale = 0.53]{am11_fig6b}\\
     {c)\hspace{2.3mm}\vspace{0mm}}\includegraphics[scale = 0.53]{am11_fig6c}
\fi 
\par\end{centering}
\caption{(Color online) Asymptotic vortex shape. The different curves are for the equilibrium shape at 
$\omega=\newomegac$, for two temperatures ($T=0.25 \tc$ and  $0.85 \tc$) at a high rotation velocity 
($\omega=452.9 \newomegac$) and for the equilibrium shape in the local induction approximation (LIA) \cite{sonin11}. 
It is expected that the same equilibrium shape as at $\omega=\newomegac$ is reached also in the low 
temperature limit $T\rightarrow 0$. a) $r$ against $z$ showing all four curves both with a unit aspect 
ratio (horizontal scale at top) and with a compressed $z$-scale (horizontal scale at bottom);
b) $\phi$ against $z$ (horizontal scale at top) and $\phi$ against $r$ (horizontal scale at bottom).
Out of the four cases, only one ($452.9 \newomegac, 0.25 \tc$) shows significant deviation from planar 
shape. In this figure $z=0$ corresponds to the end of the vortex at $r=R$. The temperatures are approximative 
values for \hethreeb, corresponding $\rea$ parameters in Table \ref{t.table}. (The arrows indicate the scale 
corresponding to the curves.)
\label{fig:Asymptotic-vortices}}
\end{figure}

These results for the full Biot-Savart model can be compared with those for a local induction 
approximation\cite{darios06,armshama65,schwarz85} (LIA) method, where the Biot-Savart integral 
(\ref{eq:v biot-savart-1}) is replaced by a local term:
\begin{equation}
\vov = \frac{\kappa}{4\pi}\ln\!\left(\frac{8}{\mathrm{e}^{1/2} \corer |\svpp|}\right) \svhp \times \svpp.
\label{eq:LIA}
\end{equation}
We have made numerical tests using this approximation. The main difference to the results above is that 
the effective value of $\Omegac$ is increased by 5\ldots7 per cent. An alternative form of LIA is to assume 
that the logarithmic factor in (\ref{eq:LIA}) is a constant, which then can be adjusted to reproduce the 
exact value of $\Omegac$ \cite{sonin11}. Even if the LIA model gives a suitable approximation for the 
velocities, it  gives wrong results in some cases, an example being the prediction $\Omegacbl=0$.

\begin{figure}[tbh]
\centerline{
\includegraphics[width=0.98\linewidth]{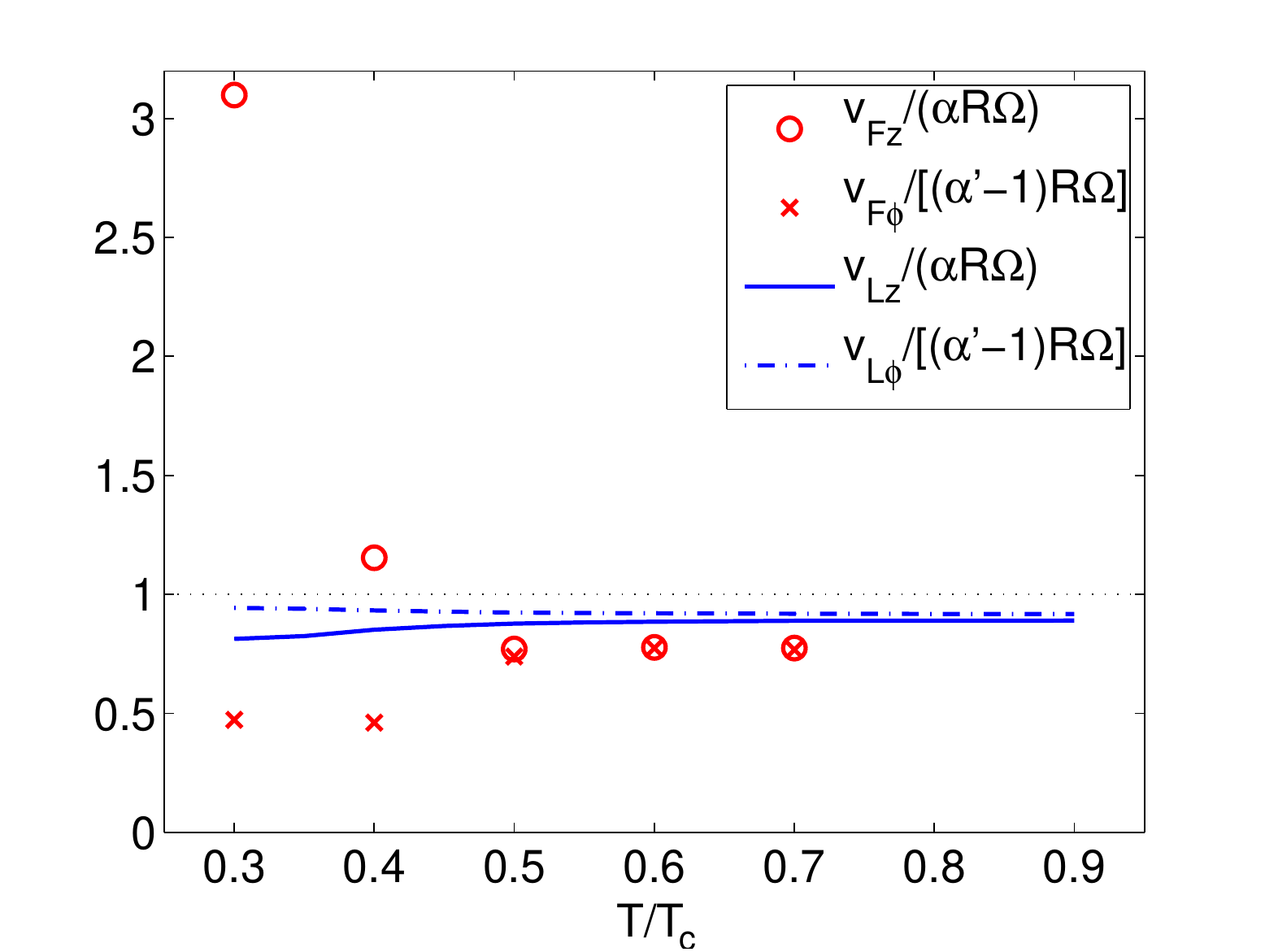} 
}
\caption{(Color online) Comparison of numerically calculated asymptotic velocities between a single 
vortex end ($\vlz$ and $\vlfi$) and a vortex front ($\vfz$ and $\vffi$). The lines are for a single 
vortex, giving the axial (solid line) and azimuthal (dash-dotted line) components. The data points are 
for the vortex front, giving the axial (circles) and the azimuthal velocities (crosses). All azimuthal 
velocities are in the rotating coordinate system. The front velocity is from numerical simulations, 
originally presented in Ref.~\onlinecite{finneetal06b}, p.~3219.
\iftwopi
     The parameters are $\newomega = 213.8$
\else
     The parameters are $\newomega = 34.0$
\fi
and $\cylr / \corer = 0.88 \times 10^5$.
\label{fig:Front-Comparison}}
\end{figure}

The shape of the top part of a rotating vortex is depicted in Fig.\:\ref{fig:Asymptotic-vortices}.
The equilibrium shape at $\omega=\newomegac$ (solid line) is independent of the friction parameters, 
as discussed above. The equilibrium shape in the local induction approximation can be solved analytically 
if a constant cut-off radius is assumed \cite{sonin11}. This result (dotted line) slightly differs from 
the result of our numerical full Biot-Savart calculation. The most important difference in the vortex 
shape is that within the LIA model the vortex approaches the rotation axis exponentially, while the 
full Biot-Savart model gives slower convergence towards the rotation axis (possibly with some power
law). This is due to vortex segments near $r\approx\cylr$, which induce an azimuthal velocity field 
that vanishes slower than exponentially. Both of these structures lie in a single plane, i.e.\ $\phi$ 
as a function of $z$ (or $r$) is a constant. At higher rotational velocity the shape of the vortex changes.
In Fig.\:\ref{fig:Asymptotic-vortices} the curve at $0.85\tc$ (dashed line) represents the dissipation-dominated 
case. In this limit the vortex is again confined to a plane. In the cross-over regime there are deviations 
from a plane, the curve at $0.25\tc$ (dash-dotted line) representing an extreme example.

The vortex shape can also be analyzed as follows. Once we know the line velocity from Figs.~\ref{fig:lambda-z} 
and \ref{fig:lambda-phi}, we can invert the equation of motion (\ref{eq: equmotion2}) to find the transverse 
part of the superfluid velocity $\vsv$ at the vortex line. We apply this to the end point of the vortex 
at the cylindrical wall ($r=R$), and find:
\begin{align}
\label{eq: v_stors}
 & \velocity_{\mathrm{s},z} = \frac{- \lambda_z + \lambda_\phi}{\rea^{-1}+\rea}\cylr(\Omega-\Omegac), \\
\label{eq: v_splane}
 & \velocity_{\mathrm{s}, \phi} = \left(\frac{\rea^{-1}\lambda_z + \rea\lambda_\phi}{\rea^{-1}+\rea}-1\right)\cylr(\Omega-\Omegac).
\end{align}
A non-zero $\velocity_{\mathrm{s},z}$ can arise from the Biot-Savart integral (\ref{eq:v biot-savart-1}) only 
if the vortex is not in a single plane. Thus the vortex can be planar only when the right-hand-side of 
Eq.\:(\ref{eq: v_stors}) vanishes. This happens only in special cases such as $\Omega=\Omegac$, or 
$T\rightarrow \tc$, or $T\rightarrow 0$, or $\omega\rightarrow \infty$.

The results above can be compared to the case of many vortices forming a vortex front. The comparison of 
the axial velocity is made in Fig.\:\ref{fig:Front-Comparison}. The velocity of a moving front is affected 
by the twisting of vortices behind the front. An analytic formula has been applied to analyze the effect of 
this twisting on the front velocity \cite{eltsovetal06,eltsovetal08}. What is studied in this paper could 
be termed as the single vortex contribution to the vortex front velocity; twisting of vortices adds another 
contribution, surface friction yet another, and phenomena associated with turbulence (reconnections, Kelvin 
cascade, vortex tangle diffusion), 
yet another. Understanding the motion of such a vortex front is an open problem. In particular, experiments 
seem to indicate that a vortex front has a nonzero propagation velocity even in the zero-temperature limit. 
This cannot be fully explained by our filament model, which lacks a mechanism of dissipating free energy at 
zero temperature.

However, such a mechanism that changes the free energy in the low temperature limit (a necessity to sustain 
the propagation of a vortex, or of a vortex front) may be provided by the pinning of vortices to the container 
wall. This pinning can be seen as an effective surface friction, pumping energy into, or out of, the system.
Also, the quasiparticle states in the vortex cores may transfer energy between the fluid and the container, when
the vortex is attached to the wall.

\section{Stability of vortices}\label{s.stability}
 
\begin{figure}[tb]
\centerline{
\includegraphics[width=0.99\linewidth]{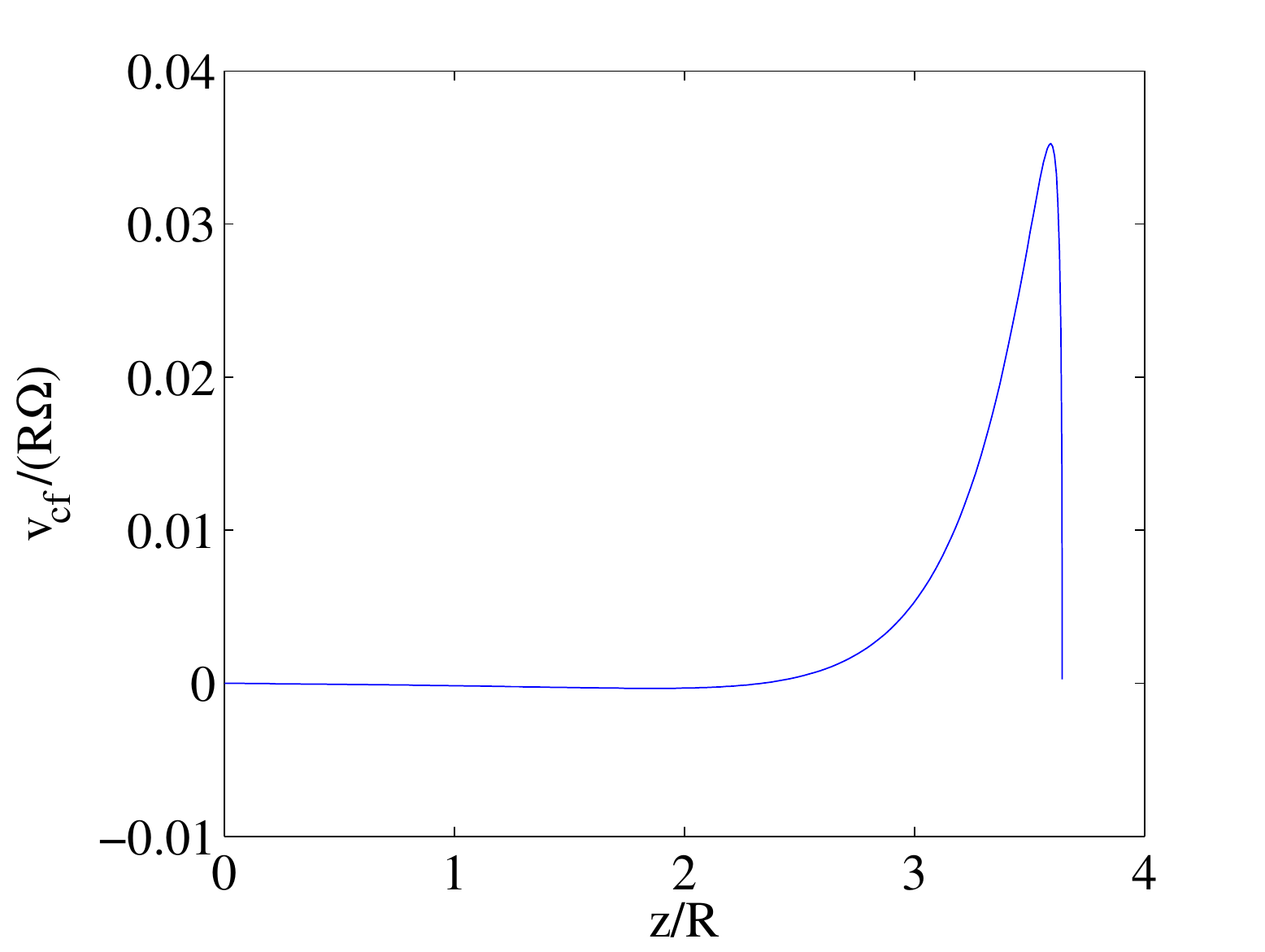}
}
\caption{(Color online) Value of counterflow projected on the vortex tangent, non-tilted cylinder. 
There is a clear peak near the top of the vortex, then a rapid drop to zero.
The critical value of projected counterflow velocity for Kelvin waves with
wavelength $\cylr$ is $\vcog = 0.1634\,\cylr\Omega$, which is clearly not reached.
The parameters are the same as for the last vortex state
in Fig.\:\ref{fig:Evolution-of-a-vortex-2}a.
\label{fig:counterflow}}
\end{figure}

Our numerical calculations indicate that once we have a vortex, whose end moves upwards, it is found to 
be stable at all temperatures, even in the zero temperature limit.
This is somewhat surprising, since the previous simulations and experimental studies in \hethreebs have 
indicated that below a certain temperature, roughly $0.5 \tc$ (or $\rea \gtrsim 1.5$), a single vortex 
becomes unstable \cite{finneetal03,finneetal06,eltsovetal09}. We suggest that this kind of single vortex 
instability is due to some surface effects (such as pinning), or heat leaks causing extra counterflow.
Earlier simulations, which also neglected surface friction, additionally assumed a cubical container instead 
of a cylindrical one \cite{finneetal03}. Studies of a decaying vortex array (spin-down) \cite{eltsovetal10}, 
also indicate that vortices in a cylindrical container are much more stable than vortices in a cubical one.

\begin{figure}[tb]
\centerline{
\includegraphics[width=0.85\linewidth]{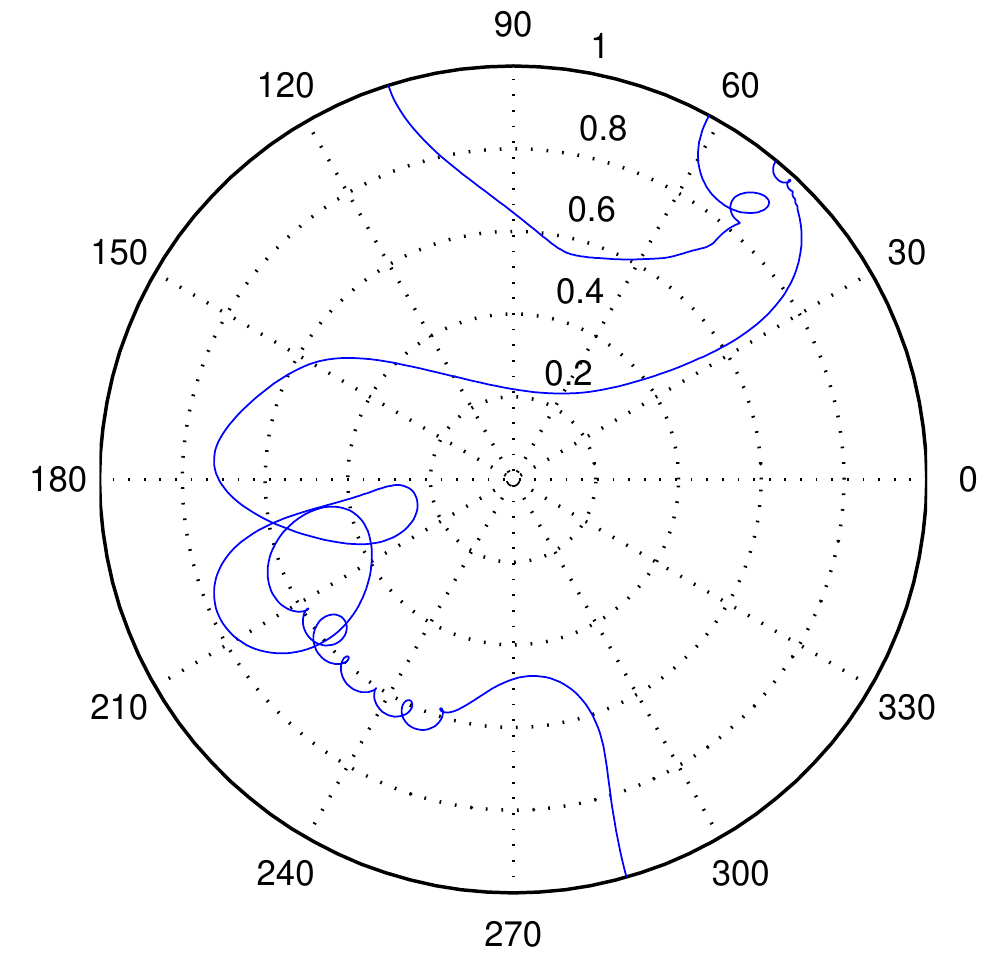} 
}
\caption{(Color online) Kelvin waves created in a rotating tilted cylinder. The appearance of Kelvin waves, 
(which can be clearly seen in the lower left quadrant of the polar plot) is due to Ostermeier-Glaberson
instability caused by the background velocity field $\vbv$. This background velocity field is nonzero in a 
rotating tilted cylinder and in this case causes a sufficiently strong counterflow along the vortex line to 
create Kelvin waves. The parameters are $\alpha\Omega t = 0.56$, $\rea = 20.6$ ($ \sim 0.3\tc$ in 
\hethreeb), $2\pi \cylr^2\Omega / \kappa = 85.5$, and $\theta = 60^{\circ}$.
\label{fig:Kelvin-waves-created}}
\end{figure}

Another source of instability is related to the creation and growth of Kelvin waves. The so-called 
Oster\-meier-Glaber\-son instability \cite{glabersonetal74,ostermeierglaberson75} appears when the 
counterflow $\vvcf := \vnv - \vsv$ along the vortex line is sufficiently strong, and causes straight vortex 
lines to become unstable towards the appearance of Kelvin waves. For a single axially oriented straight 
vortex, the critical counterflow velocity, above which the amplitude of Kelvin waves with a wave vector 
$k$ are able to grow, is:
\begin{equation}
\vcog = \frac{\Omega+\eta k^{2}}{k}, \textrm{ where }
\eta = \frac{\kappa}{4\pi} \ln\!\frac{1}{k\corer}.
\label{eq:ostglab}
\end{equation}
In our case of a non-tilted cylinder with a vortex attached to the cylinder wall this effect may exist in some 
situations. From Fig.\:\ref{fig:Evolution-of-a-vortex-2}b one observes that the vortex has a small component 
along the azimuthal direction. This results in a non-zero counterflow along the vortex and could in principle 
cause Ostermeier-Glaberson instability. 

However, growing Kelvin waves are not observed in simulations when the tilt is absent. The reason becomes 
apparent if one looks at Fig.\:\ref{fig:counterflow}, where we have plotted the counterflow along the 
vortex. The counterflow along the vortex line has its maximum in the top part of the vortex and falls 
sharply below it. Near the axis it is practically zero. The maximum value for the counterflow indicates 
that the Kelvin waves that could grow have a wavelength much longer than the cylinder radius, and, therefore, 
larger than the region of this finite counterflow. Hence, the Kelvin waves cannot grow and Ostermeier-Glaberson 
instability does not play any role. 

In contrast, the situation in a tilted rotating cylinder is different. With enough tilting, the boundary 
superfluid velocity $\vbv$ creates a sufficiently strong counterflow, which results in the appearance
of Kelvin waves. These Kelvin waves then produce a change from a laminar to a more turbulent fluid motion, 
possibly resulting in a vortex tangle, see Fig.\:\ref{fig:Kelvin-waves-created}.
Yet another phenomenon is the Crow instability \cite{crow70} which applies to vortices that are close to each other, 
or close to the wall (which can be interpreted as being close to the image vortex). This instability is due to the fact 
that vortices of opposite direction attract each other. Thus, the parts of vortices that are slightly closer to each 
other start moving faster towards each other, deforming the vortices even more, and eventually resulting in a large 
number of reconnections and new vortices. In our case the Crow instability does not work in the untilted cylinder 
since the rotation provides a stabilizing effect.

\begin{figure}[tb]
\includegraphics[width=0.99\linewidth]{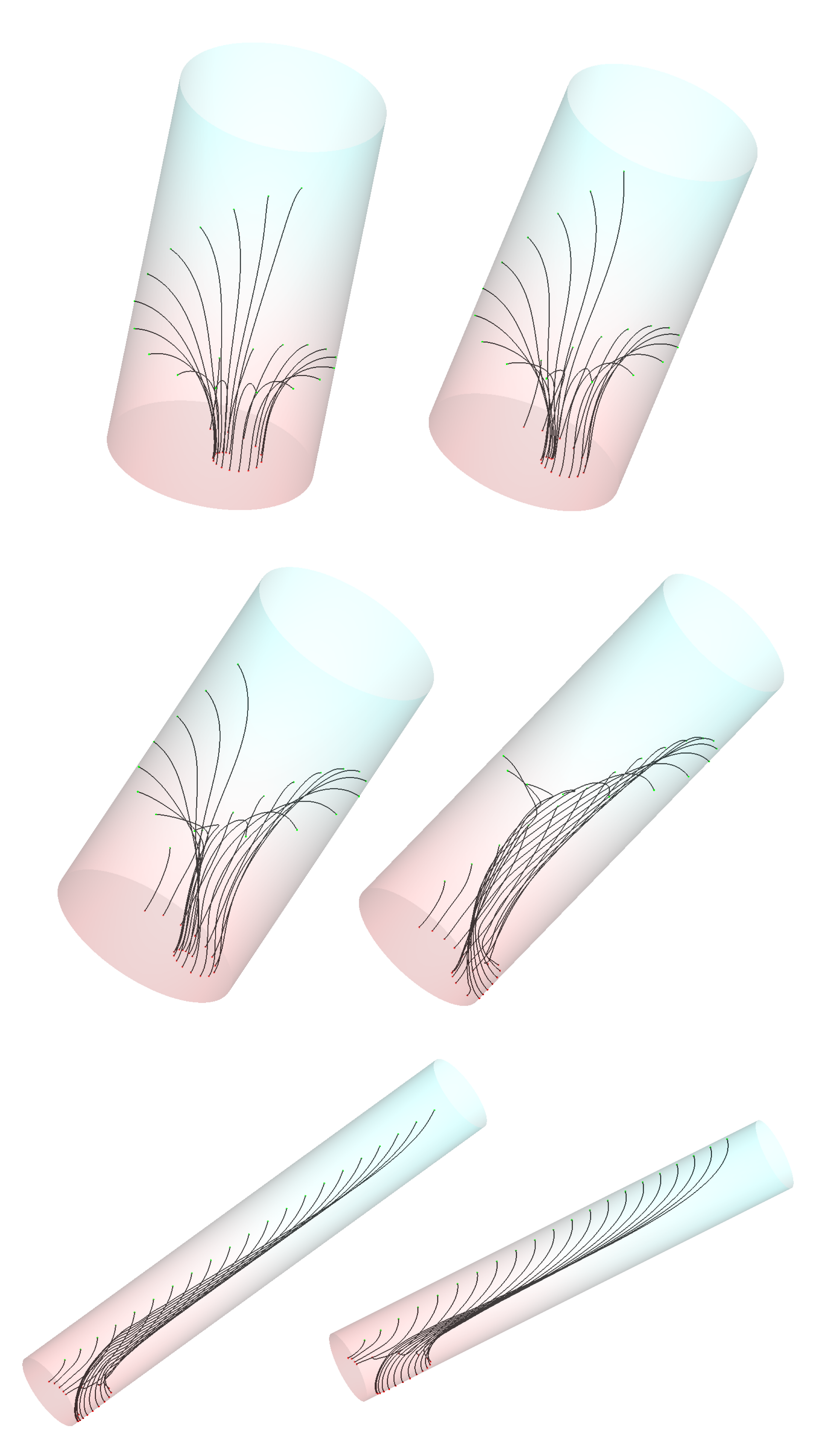}
\caption{(Color online) Vortex evolution for different tilting angles. Twenty-one snapshots 
are shown for  each tilting  with angles $\theta = 10^{\circ}$, $20^{\circ}$, $30^{\circ}$, $40^{\circ}$, 
$50^{\circ}$, and $60^{\circ}$. With increasing tilting some qualitative differences appear. The 
starting point of the vortex may climb on the side wall. The end point may start moving axially.
All plots are in the rotating coordinate system. The total time is $2.33 / \alpha\Omega$,
$\rea = 3.63$, and $\newomega = 85.5$.
\label{fig:Effects-of-tilting}}
\end{figure}

\section{Effect of tilting}\label{s.tilt}

Tilting the vessel has the advantage of breaking the cylindrical symmetry. A small tilt may be useful 
in calculations also for detecting effects that occur due to this asymmetry, unavoidably present in 
experimental situations.

The most prominent effect caused by the tilt is, as noted above, the instability of the vortex. At low 
temperatures and for a large tilt, several initial configurations (at large enough rotation) lead to growing 
Kelvin waves and eventually to a creation of a vortex tangle via reconnections. However, with an initial state 
close to the steady state, the vortex smoothly adopts its new asymptotic form and no new vortices are generated.
  
A second noticeable effect of tilting is the oscillation of $\vlz$ about a constant value (close to 
$\alpha\cylr\Omega\cos\theta$). This is due to a sinusoidal component of the boundary field \cite{hanninen09}.

A somewhat surprising effect was also found. At a large enough tilt angle the azimuthal velocity of the vortex 
end may approach zero (in moving coordinates). Thus, in the asymptotic limit, the vortex end becomes locked at 
some azimuthal angle, while a constant axial velocity component remains. A detailed study of this phenomenon has
not been done yet, but this sort of behavior can be seen in the last two pictures of Fig.\:\ref{fig:Effects-of-tilting}, 
corresponding to tilting angles $50^{\circ}$ and $60^{\circ}$.

\section{Conclusions}\label{s.conc}

While one could say that we have been ``hitting a mosquito with a cannonball'', using a code created for brute 
force large scale vortex tangle calculations to do a very simple job, there still is, we believe, some potentially 
useful information to be gained from this endeavor.

Vortex motion in a cylinder with smooth walls was found to be quite stable, both in the untilted and moderately 
tilted ($\lesssim 30^{\circ}$) cases. This limits the causes for the experimentally observed single vortex instability
\cite{finneetal03,finneetal06,eltsovetal09} to areas not covered in this study, such as surface effects and heat flows.

In our ideal cylindrical environment, vortex motion can be estimated by considering the limiting cases at 
$T\rightarrow 0$, $\Omega = \Omegac$, and $\Omega \rightarrow \infty$. In a more general case the correct motion 
can be characterized by introducing a small correction to these limiting cases. Our scaling argument additionally 
emphasizes that this correction depends only on the dimensionless parameters $\cylr^2\Omega/\kappa$, $\rea$, and 
$\cylr/\corer$, the last dependence being weak (logarithmic). At large enough rotational velocities the vortex 
motion is dominated by {\em dissipative} effects. Near $\Omegac$ one may observe an {\em inertial regime}  
where non-dissipative forces dominate. In general, the asymptotic vortex shape is three-dimensional, but e.g., 
at $\Omega = \Omegac$ the vortex configuration is confined to a plane. Deviations from the plane structure are 
typically small, the largest deviations appearing at low temperatures, and with relatively large rotation 
velocities.  

This study highlights the role of scaling properties in superfluids and other 
analogous systems. A deeper analysis to connect these formulas to the velocity of 
the vortex front should be carried out. The azimuthally locked motion in the tilted 
case was an unexpected result.

\acknowledgments
We would like to express our gratitude to Professor Alexander L. Fetter for
useful comments and discussions.
This work is supported by the Academy of Finland 
and EU 7th Framework Programme (FP7/2007-2013, grant 228464 Microkelvin).

\bibliography{am11}

\end{document}